# Through-chip microchannels for three-dimensional integrated circuits cooling

Lihong Ao and Aymeric Ramiere*

College of Physics and Optoelectronic Engineering, Shenzhen University, Shenzhen 518000, P. R. China

*Corresponding author. E-mail: ramiere@szu.edu.cn

**Abstract:**

Cooling high-power electronics in multilayer integrated circuits (ICs) is challenging for existing cooling methods. In this work, we designed through-chip microchannels (TCMCs) that cross the entire chip perpendicularly to the layers, with water circulating inside to provide direct cooling to each layer. TCMCs are organized in a square array where the pitch and radius of the microchannels are explored. Our computational fluid dynamics (CFD) simulations show that a pitch ~10 μm and a radius ~1 μm optimize the cooling performance to support a power higher than $10^4$ W/cm² while the maximum temperature rise remains below 60 K with a water inlet temperature of 300 K. We show that the cooling properties do not change with the number of layers for a given chip thickness which provides flexibility to the functional design of the chip. Though manufacturing may be challenging, TCMCs offer a new way for chip cooling that could provide a leap forward in the performance of multilayer 3D ICs and high-power electronics.

Keywords: thermal management; liquid cooling; microchannels; three-dimensional integrated circuits; through-chip;

**Nomenclature**

| | |
|---|---|
| $R$ | Radius of microchannels (m) |
| $a$ | Distance between microchannels (m) |
| $L_{MC}$ | Length of microchannels (m) |
| $N$ | Number of layers |
| $\mathbf{n}$ | Unit vector |
| $P$ | Pressure (Pa) |
| $P_{in}$ | Inlet pressure (Pa) |
| $\mathbf{u}$ | Velocity (m/s) |
| $\mathbf{u}_w$ | Slip velocity on walls (m/s) |
| $\boldsymbol{\tau}_{n,t}$ | Tangential wall stress (Pa) |
| $Q_h$ | Heat source power density (W/m²) |
| $q_h$ | Heat source volume power density (W/m³) |
| $Q_{\text{fric}}$ | Frictional power density (W/m²) |
| $Q_{\text{pump}}$ | Fluid pumping power density (W/m²) |
| $q_f$ | Fluid volume power density (W/m³) |
| $\mathbf{q}$ | Heat flux density (W/m²) |
| $T$ | Temperature (K) |
| $T_{\max}$ | Maximum temperature (K) |
| $T_{in}$ | Inlet temperature (K) |
| $\Delta T_{max}$ | Maximum temperature difference (K) |
| $\langle \Delta T_w \rangle$ | Average temperature rise at the outlet (K) |
| $C_p$ | Specific heat of water (J/kg/K) |
| $\alpha_v$ | Tangential accommodation momentum coefficient |
| $\lambda_f$ | Mean free path of water (m) |
| $\mu_f$ | Viscosity of water (Pa.s) |
| $\rho_f$ | Density of water (kg/m3) |
| $\kappa_f$ | Thermal conductivity of water (W/m/K) |
| $\phi_f$ | Mass flux density of water (kg/s/cm²) |

## 1. Introduction

Three-dimensional integrated circuits (3D ICs), or multilayer semiconductor devices, involve stacking multiple layers of integrated circuits on top of one another. 3D ICs integrate various components onto a single chip, which increases functionality, improves performance, reduces size, and increases the power efficiency of electronic devices [1,2]. 3D ICs are also crucial for increasing electronic density by stacking IC layers, as flat chip processing is slow and incremental today. Thermal management is one of the most challenging aspects of developing 3D ICs because of the high thermal power density and vertical thermal coupling between the components [3–7]. Various 3D integration strategies are being explored, which hold great expectations for future electronic applications such as high-performance computing, quantum computing, telecommunications, and high-power chips [8–10]. However, the performance of 3D ICs is restrained by thermal issues today [11]. Therefore, a performant cooling approach is highly needed.

Fluid cooling with microchannel heat sinks (MCHSs) is one of the most promising strategies for 3D IC cooling. It has been intensively studied over the past few years thanks to the universalization of nanofabrication techniques and 3D printing [12–14]. MCHSs for single-phase fluids appear in many configurations and offer many possibilities for optimization [13–15]. Microchannels can be straight [16], straight with interior decorations [17], wavy [18], porous fins [19], or have more complex designs [20–22]. MCHSs also appear as arrays of pins with various sizes and arrangements [23–27]. Hybrid designs combining microchannels and pins or fins have also been recently proposed [28–31]. Today, after many optimizations, the cooling power density of MCHSs is around 1000 W/cm$^2$, with the best-recorded MCHSs design that reaches a cooling power density of 1700 W/cm$^2$ [32]. However, future 3D ICs would require considerable cooling power to increase computing performance and functionalization.

MCHSs are usually integrated as independent devices placed on top of the chip, which provides limited cooling to the deep layers of 3D ICs. Lu and Vafai proposed a design of counterflow MCHSs placed on both the chip's top and bottom, demonstrating good performance for a three-layer chip [33]. Recently, Huang et al. extended this concept up to five layers of microchannels and optimized the design [34]. However, the cooling still originates from only one side of the chip. The performance is expected to decrease when more IC layers are included as the heat in the middle becomes increasingly difficult to remove. Finally, Xia *et al*. proposed intercalating microchannel system between each layer of the 3D ICs [35]. Such a design is thermally effective but raises fabrication challenges and problems with the vertical integration of the layers. Therefore, new solutions are necessary to enable 3D ICs with more than three layers.

Besides MCHSs, another way to manage heat in 3D ICs can be achieved by the through-silicon vias (TSVs) that interconnect the different levels of layers electrically, forming a network that allows the layers to work as one chip [36,37]. Though TSVs were not initially intended for heat management, their positioning can be optimized to increase heat dissipation and avoid hot spot formation [38,39]. The drilling technology developed for TSVs is mature today and allows obtaining relatively smooth and parallel walls through silicon layers of more than 100 µm [40]. Oh et al. experimentally demonstrated that it is possible to integrate a microfluidic system within the TSVs [41]. Wang et al. recently proposed to use TSVs in complement with MCHSs to provide more efficient cooling of 3D ICs, but still separating the MCHSs from the IC so that the cooling power comes mainly from the bottom of the chip [42]. The prospect of fully combining the technology of TSVs with the cooling performance of MCHSs is very interesting for high-performance cooling of 3D ICs, but no design has been proposed yet.

In this paper, we propose through-chip microchannels (TCMCs) consisting of an array of microchannels (MCs) perpendicular to the layers and cross through the entire chip. Contrary to current technology, where the IC and cooling system are separated, TCMCs intertwine these two parts to provide direct cooling to every layer in the 3D IC. Computational fluid dynamics (CFD) simulations are conducted to determine the thermal performance of the design. The TCMCs are arranged in a square array where the MCs' pitch and radius are investigated. The impact of the device power, the inlet pressure, and the number of layers are also studied. Our results show that a radius of 1.5 µm minimizes the temperature increase. In these conditions, a maximum power of 15 kW.cm$^{-2}$ can be applied before reaching a temperature increase of 60 K, improving the cooling power density by one order of magnitude compared with traditional MC designs. Furthermore, for a given chip thickness, we find that the number of layers does not impact the thermal performance when the power is distributed among the layers. Despite potential fabrication difficulties, the exceptionally high cooling performance of TCMCs could represent a breakthrough for future high-performance electronic devices.

## 2. Numerical methods
### 2.1. TCMC Design

The TCMC design consists of a square array of microchannels that cross through the whole thickness of the chip perpendicularly to the layers. Fig. 1(a) shows the unit cell used for the CFD simulations. The microchannels have a pitch $a$ that is varied from 4 µm to 30 µm and a radius $R$ that is varied from 0.6 µm to 3.8 µm. Water is chosen as the coolant flowing in the MC. The MC length is also the total thickness of the domain that is maintained constant at $L_{MC} = 37.5$ µm. A number N of layers from 1 to 5 is distributed within this thickness. Each layer comprises three sub-layers: a basis, a heat source, and an insulation, as shown in Fig. 1(b). The basis consists of silicon separating the layers on the chip. The heat source is a thin film simulating the heat dissipated by the transistors. A total heating power $Q_h$ is distributed equally among the N heating sub-layers and propagates by conduction into the solid layers. The insulation is set to SiO$_2$, which has much lower thermal conductance than silicon, to simulate the heat transfer obstruction between the layers. The heat source has a constant thickness $\delta = 1$ µm while the basis and insulation adapt to represent 85% and 15% of the remaining layer thickness, respectively.

The TCMC design has several advantages compared to MCHS. First, the through-silicon design provides direct cooling of every layer, even those in the middle of the chip, which could increase the number of layers that can be stacked. Second, the regular micrometer-spaced microchannels create short-distance cooling everywhere in the chip, limiting the formation of hot spots. Third, despite the micrometer scale of the MC, the pressure drop is not significant, thanks to the short length of the microchannels. Finally, the whole surface area of the microchannels is in contact with the layers to provide a cooling area higher than the surface under the chip.

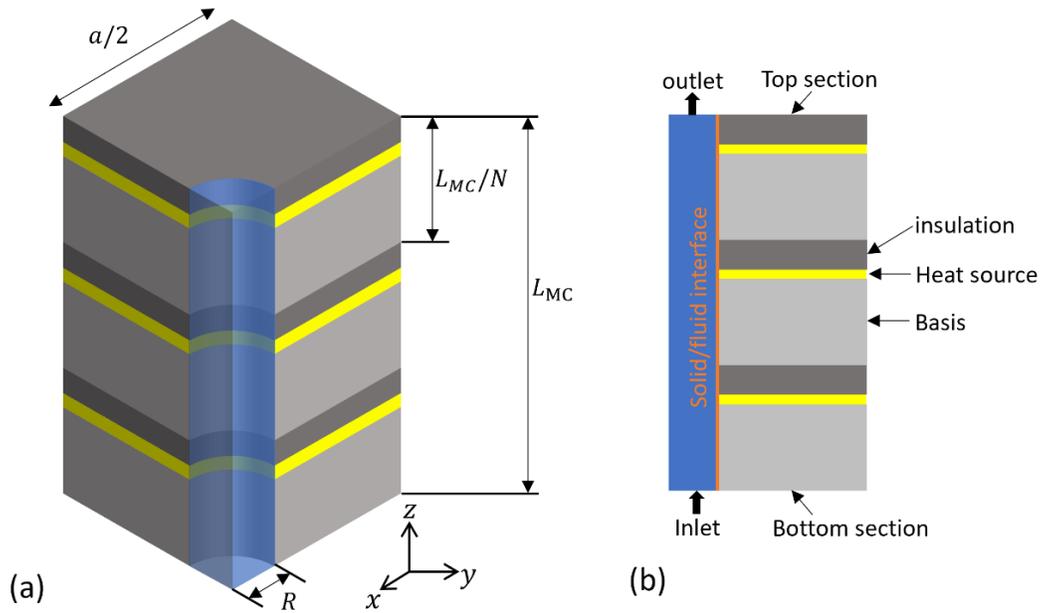

**Fig. 1.** Unit cell design of TCMC. (a) 3D view. (b) side view.

Our CFD simulations aim to evaluate the TCMC design's performance to limit the temperature increase within the functional layers of a potential 3D IC. The substrate below the functional layers is ignored because the aperture of this substrate is one order of magnitude larger than $R$, meaning that almost all heat transfer and pressure drop occur in microchannels.

### 2.2. Numerical modeling

CFD simulations were performed using the finite element method using Comsol Multiphysics v.6 software. Symmetrical boundary conditions are applied on the four sides of the unit cell, as shown in Fig. 1(a). Hexahedral meshes for both the solid and liquid domains were used. Mesh conditions were verified to obtain good accuracy in all the cases (see Appendix A). The steady-state heat transfer is calculated in the whole domain and the water flow in the MC, with particular attention to the solid/fluid interface. The equations and boundary conditions are given below.

The heat conduction equation governs the heat transfer in the solid domain:

$$-\kappa_i \nabla^2 T = q_i, \tag{1}$$

where $\kappa_i$ is the thermal conductivity and $q_i$ the source term for the sublayer $i$. The thermal conductivity in silicon is supposed to be constant with $\kappa_{si} = 130\ W.m^{-1}.K^{-1}$, while it is temperature-dependent in $SiO_2$ (see Appendix B). No heat source is located in the basis and insulation sub-layers such that $q_{bas} = 0$ and $q_{ins} = 0$. For the heating sub-layer, $\kappa_h = \kappa_{si}$ and the source term is defined as:

$$q_h = \frac{Q_h}{N\delta} \frac{a^2}{a^2 - \pi R^2}, \tag{2}$$

where $Q_h$ is the surface power density, $N$ is the number of layers, and $\delta = 1\ \mu m$ is the thickness of the heating sub-layer. A power of $Q_h a^2/(4N)$ is applied to a layer of the computational domain. Perfect contact between the solid layers is assumed such that there is no temperature jump at the solid/solid interfaces. The top and bottom sections of the solid domain are modelized with adiabatic conditions such that the heat flux at these surfaces is $q = 0$ (see Fig. 1(b)). These conditions are expected to yield higher temperatures than in reality, where the surrounding environment can remove some heat.

In the fluid domain, Generalized Navier-Stokes Equations (GNSE) in steady state and Maxwell slip wall boundary conditions are applied to model the motion and temperature of the water. The water flow is supposed to be near-incompressible and laminar. In all our simulations, the MC radius is above 1 µm, which is much bigger than the mean free path of water ($\sim 10^{-4}\ \mu m$), ensuring the applicability of GNSE. Within the fluid, the GNSE equations read:

$$\nabla \cdot (\rho_f \boldsymbol{u}) = 0 \tag{3}$$

$$\rho_f(\boldsymbol{u} \cdot \nabla)\boldsymbol{u} = \nabla \cdot \left(-p\boldsymbol{I} + \mu_f(\nabla\boldsymbol{u} + (\nabla\boldsymbol{u})^T) - \frac{2}{3}\mu_f(\nabla \cdot \boldsymbol{u})\boldsymbol{I}\right) \tag{4}$$

$$\rho_f C_p \boldsymbol{u} \cdot \nabla T - \kappa_f \nabla^2 T = q_f \tag{5}$$

where Eq. (3), Eq. (4), and Eq. (5) are the continuity mass equation, the momentum equation, and the energy equation, respectively. $\rho_f, \mu_f, C_p, \kappa_f$ are temperature-dependent density, viscosity, specific heat, and thermal conductivity of water [43]. More information about the thermophysical properties is given in Table 1 of Appendix B. $\boldsymbol{I}$ is a 3x3 unit matrix. The source term in Eq. 5 is $q_f = Q_{fric}/(\pi R^2 L_{MC})$, with $L_{MC}$ the MC length. The term is due to the fluid flow that generates heat by friction in the MC following the equation:

$$Q_{\text{fric}} = \int_{\text{out}} dS\, P_{\text{in}} u_z \tag{6}$$

where $P_{in}$ is the water inlet pressure. $Q_{\text{fric}}$ was simulated independently for different inlet pressures and MC diameters. The obtained data were fitted as a function of $R$, $L_{MC}$, and $P_{in}$ by a set of external simulations. The fit function of $Q_{fric}$ was used in Eq. 5 as a known parameter (see Appendix B).

Inlet condition includes pressure inlet condition and heat inlet condition. We use the inlet pressure instead of the inlet velocity because the pressure can be directly controlled via external devices. Heat inlet condition infers that temperature change of inflow fluid causes a heat flux through the inlet boundary, that is

$$\boldsymbol{q} \cdot \boldsymbol{n} = -\int_{T_{\text{in}}}^{T} C_p\, dT \rho_f(T) \boldsymbol{u} \cdot \boldsymbol{n} \tag{7}$$

where $\boldsymbol{q}$ is the heat flux at the inlet boundary, and $\boldsymbol{n}$ is the normal unit vector. $T_{\text{in}} = 300K$ is the initial temperature of the coolant, $T$ is the temperature field.

The solid/fluid interface is critical to the heat transfer. Maxwell slip wall boundary conditions were applied as Churaev *et al.* indicated that the slippage effect significantly impacts fluid flow in lyophobic capillaries with diameters below a few micrometers [44]. In Maxwell slip wall boundary conditions, velocity slip and temperature jump on the wall can be expressed as [45]

$$\boldsymbol{u} - \boldsymbol{u}_w = \frac{2-\alpha_v}{\alpha_v}\frac{\lambda_f}{\mu}\boldsymbol{\tau}_{n,t} + 0.75\frac{\mu_f}{\rho_f T_f}\boldsymbol{\nabla}_t T_f \tag{8}$$

$$T_w = T_f - \frac{2-\alpha_v}{\alpha_v}\frac{2\gamma}{1+\gamma}\frac{\kappa_f \lambda_f}{\mu_f C_p}\boldsymbol{n} \cdot \boldsymbol{\nabla} T_f \tag{9}$$

where $\boldsymbol{u} - \boldsymbol{u}_w$ is the slip velocity, $T_w$ is wall temperature, and $T_f$ fluid temperature. $\lambda_f$, $\boldsymbol{\tau}_{n,t}$, $\boldsymbol{\nabla}_t T_f$ are the mean free path, the tangential component of wall stress, and the temperature gradient at the local wall, respectively. $\gamma$ is the ratio of specific heat. $\alpha_v = 0.9$ is the momentum accommodation coefficient in Maxwell slip wall condition.

The simulation results provide the temperature distribution T in the silicon materials and water as well as the water velocity $\boldsymbol{u}$ and water pressure $p$ in the fluid domain. Four quantities are derived from these results, as described below. First, the maximum temperature difference is given by

$$\Delta T_{max} = T_{max} - T_{in} \tag{10}$$

where $T_{max}$ is the highest temperature and $T_{in} = 300$ K is the inlet temperature that is fixed. The phase change of the water coolant is not included in the simulation, so the results with $\Delta T_{max} > 73\ K$ are deleted in the results. Actually, $T_{max}$ is reached in the solid domain, while the maximum temperature in the water is smaller. The limit of 73 K is, therefore, a conservative value with a margin to the boiling water.

Finally, three quantities of the water flow are derived: the mass flux, the pumping power, and the average temperature rise. The water mass flux $\phi_f$ is determined by

$$\phi_f = \frac{1}{a^2} \int_{in} \rho_f u_z\, dS \tag{11}$$

where only the vertical direction $u_z$ of the water is accounted for, and the integral is over the surface $dS$ of the MC. The normalization is over the area under the unit cell, which gives the mass flux for the chip instead of the inlet section of the MC. Then, we calculate the water pumping power density required to overcome the viscous drag and move the water through the MC with an inlet pressure $P_{in}$:

$$Q_{\text{pump}} = \frac{1}{a^2} \int_{in} dS \left( P_{in} + \frac{\rho_f \boldsymbol{u}^2}{2} \right) u_z \tag{12}$$

Eq. 12 corresponds to the addition of the frictional power in Eq. 6 and an acceleration power based on the density and velocity of the water. $Q_{\text{pump}}$ should be below $10\ kW.cm^{-2}$ to maintain reasonable power requirements. Last, we obtain an approximate average water temperature rise using the formula:

$$\langle \Delta T_w \rangle = \frac{Q_h + Q_{\text{pump}}}{C_p(T_{in})\phi_f} - T_{in} \tag{13}$$

$\langle \Delta T_w \rangle$ corresponds to the average water temperature at the outlet of MC. In practical applications, if a circulation cooling system is used outside the chip, $\langle \Delta T_w \rangle$ should be high enough to make the outer components (such as a cooling tower) work well.

## 3. Results and discussion

### 3.1. Optimization of microchannel design

We take the case with a heating power $Q_h = 10\ kW.cm^{-2}$ and inlet fluid pressure $P_{in} = 1$ MPa as our reference to conduct a detailed study of the impact of the design parameters. Fig. 2(a) shows the temperature distribution in the steady state both in the solid and fluid domains for a three-layer chip with a pitch $a = 10\ \mu m$ and a radius $R = 1.5\ \mu m$. The water enters the MC with the temperature $T_{in} = 300$ K, providing the lowest solid temperature in the first layer. Then, the water temperature increases when moving up the MC, and the solid temperature rises consequently. The maximum temperature is reached in the top left corner of the unit cell, which is the farthest from the MC. In this case, $\Delta T_{max} = 344.7$ K. Due to heat resistance between layers, heat generated by the layers in high position is difficult to propagate to the bottom of the chip, resulting in relatively uniform temperature within each layer.

The velocity profile remains identical throughout the MC, as shown in Fig. 2(b). Laminar slip flow remains the same profile from the inlet to the outlet, with a velocity almost only with the z component. It is easy for the velocity boundary layer to extend to the center of the MC in microscales. It indicates that the pressure drop in the MC is negligible because they are very short. For that case, we obtain $\phi = 1059\ g.cm^{-2}.s^{-1}$, $Q_{pump} = 19.47\ W$, and $\langle \Delta T_w \rangle = 22.8\ K$. The water velocity at the center of the MC increases from 8.7 m/s for $P_{in} = 0.2$ MPa to 50.4 m/s for $P_{in} = 2.0$ MPa. These velocities were compared to the analytical solution of the Poiseuille flow to validate our simulations (see Appendix A).

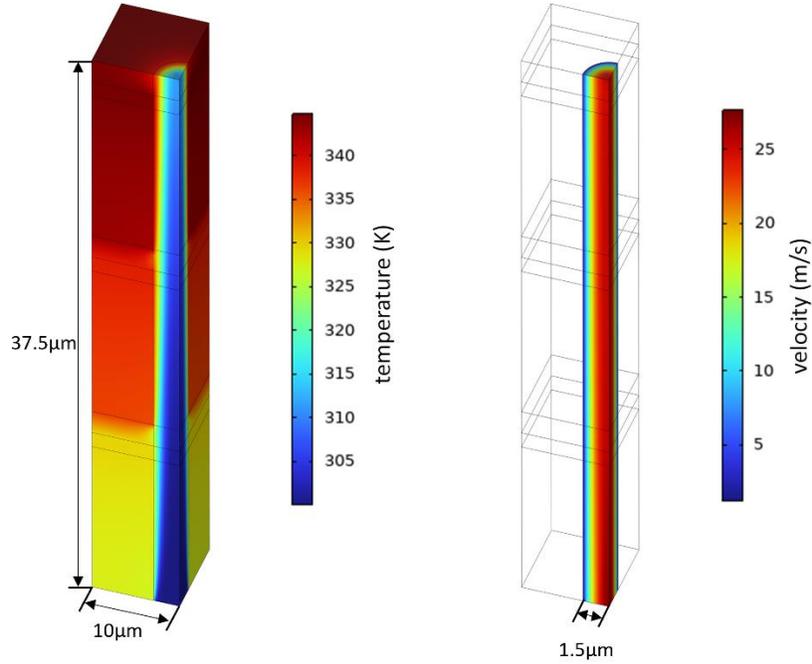

**Fig. 2.** Temperature distribution in the whole simulation domain (a) and water velocity distribution in the microchannel (b) for a three-layers chip when $a = 10 \mu m$, $R = 1.5\ \mu m$, $Q_h = 10\ kW.cm^{-2}$, $P_{in} = 1 MPa$.

In our design, the main parameters are the pitch $a$ between two MCs, and their radius $R$. The pitch-to-radius ratio $R/a$ is a more convenient parameter for scaling purposes than the radius alone. Using the data as shown in Fig. 2, we can derive the quantities given in Eq. 10 to Eq. 13 and obtain contour maps as shown in Fig. 3. The data where $\Delta T_{max} > 73$ K were removed from the plot as it exceeds the temperature of boiling water under atmospheric pressure. This is a safe limit as the pressure in the MC is higher and, therefore, can sustain liquid water to a higher temperature. The data with $Q_{pump} > 10\ kW.cm^{-2}$ were also removed as the power exceeded conventional application conditions. The maximum temperature increase $\Delta T_{max}$ is shown in Fig. 3(a). As expected, $\Delta T_{max}$ decreases as $R/a$ increases such that the limit of 73 K is reached for small $R/a$. Interestingly, an optimum value around $a = 10\ \mu m$ is found to yield low $\Delta T_{max}$. This optimum results from the competition between the distance separating MCs and the water flow within the MC. Indeed, as $a$ increases, the distance between MCs increases, and the heat in the middle becomes more difficult to remove. The sudden increase of $\Delta T_{max}$ for small $a$ follows the line of constant radius $R = 1\ \mu m$. Below this radius, the water flow is dominated by the solid/fluid boundary conditions that restrain the flow, as modelized by our slip conditions.

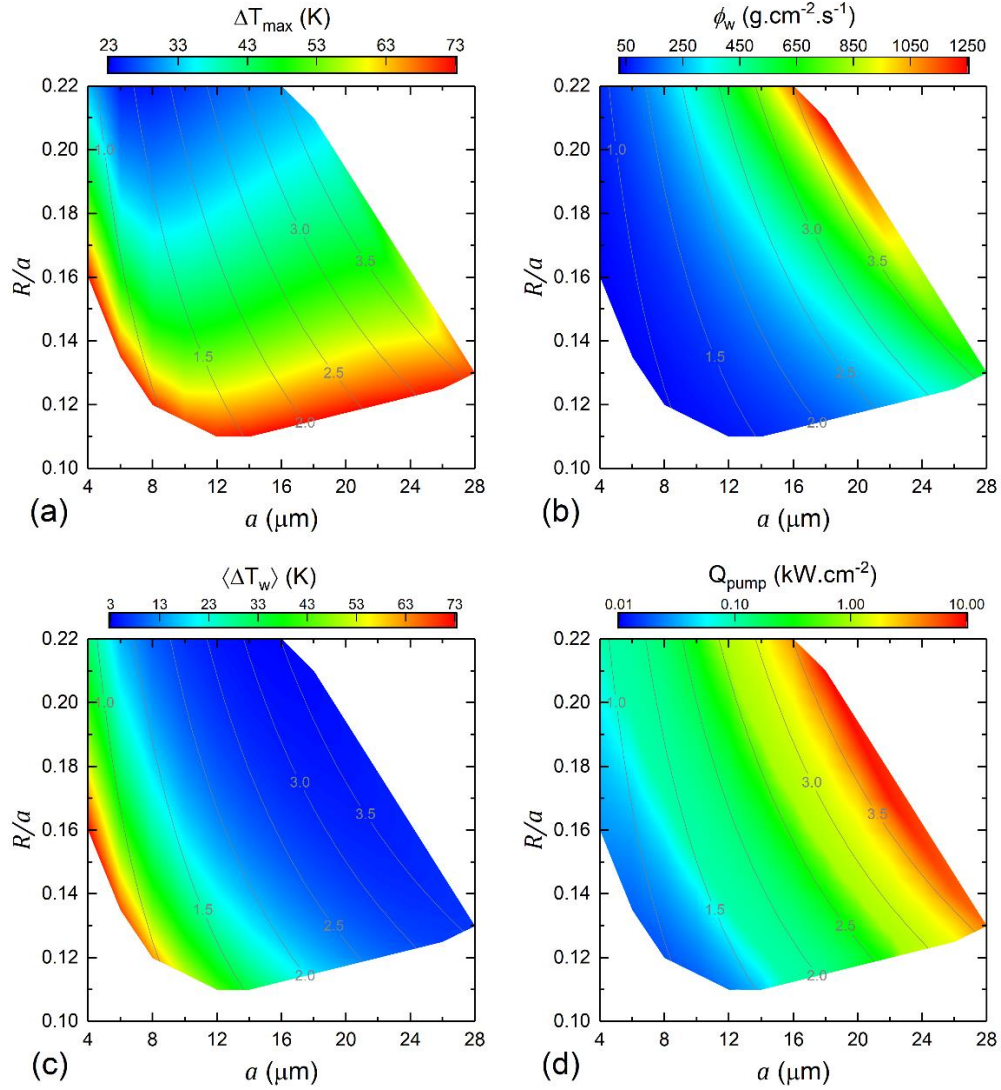

**Fig. 3.** (a) Maximum temperature difference $\Delta T_{max}$. (b) Water mass flux $\phi_w$. (c) Average water temperature rise $\langle \Delta T_w \rangle$. (d) Pumping power $Q_{pump}$. The grey lines indicate the MC radius in µm. $Q_h = 10 \ kW.cm^{-2}$ and $P_{in} = 1 \ MPa$ for all the simulations.

The diminution of the water flow $\phi_w$ is visible as the mass flux decreases quickly when $R$ decreases, as shown in Fig. 3(b). This trend is expected as we fix $P_{in}$ independently of the pitch so that $\phi_w$ is mainly determined by the MC radius. The Reynolds number Re can be estimated using $\rho R |\boldsymbol{u}|/\mu$. Given the range of water mass flux, Re lies between 60 and 1500, which confirms that we remain in laminar flow conditions. In Fig. 3(c), the average outlet water temperature rise follows approximately the MC radius. It increases more rapidly when $R < 1.5 \ \mu m$ when the slip conditions become significant. For $R > 2.0 \ \mu m$, little variations are observed, with the temperature rise generally below 10 K because the water temperature decreases rapidly from the MC walls. The pumping power $Q_{pump}$ in Fig. 3(d) increases quickly with the MC radius and was capped at $10\ kW.cm^{-2}$ to limit the power consumption. An MC radius below 2.5 µm necessitates reasonable powers of less than $1\ kW.cm^{-2}$. It shows that TCMCs pumping power consumption is within accessible range with today's industry standards.

Given the results shown in Fig. 3, further studies in this paper focus on the pitch $a = 10\ \mu m$ that minimizes $\Delta T_{max}$. This pitch also provides reasonable conditions with $Q_{pump} < 0.5\ kW.cm^{-2}$, $\phi < 250\ g.cm^{-2}.s^{-1}$ and $\langle \Delta T_w \rangle < 45\ K$.

### 3.2. Impact of heating power density $Q_h$

Here, we study the relation between the heating power density $Q_h$ and the maximum temperature in the chip $\Delta T_{max}$. We used three layers with a constant pitch $a = 10\ \mu m$ and an inlet pressure $P_{in} = 1\ MPa$. Fig. 4(a) shows that $\Delta T_{max}$ increases more rapidly with $Q_h$ when the MC radius decreases. The limit of $\Delta T_{max} = 60K$ is taken to calculate a maximum heating power $Q_{h,60K}$ that could be supported in the chip. The smallest simulated radius of $1.2\ \mu m$ can already support a heating power of $9\ kW.cm^{-2}$ and then easily exceeds $10\ kW.cm^{-2}$ thanks to the dense array of MCs that quickly removes the heat.

Fig. 4(b) displays a linear relationship within our radius variation range, following the equation $Q_{h,60K} = -10.6 + 16.5R$. The negative ordinate at the origin indicates that this linear law does not work for very small radii for which proximity effects occur due to the slip conditions at the surface of the MC walls. This effect reduces as the MC radius increases to vanish in our simulations, where we observe only a linear law that should also hold for bigger radii beyond 2 µm. While a larger radius offers better cooling power, it also means less surface to include electronics. The inset in Fig. 4(b) shows the surface ratio of the chip occupied by the MC, which is given by $\pi R^2/a^2$. Our simulations are limited to $R/a < 0.22$, leading to an MC occupation ratio of 15%. It is only 7% when $R/a = 0.15$ ($a = 10\mu m$ and $R = 1.5\mu m$). Therefore, our TCMC design provides ample room to include electronics. Furthermore, the distance between the holes is large enough to include a connection path to ensure that every component in the IC can be interconnected.

The simulated cooling performance is excellent since even a radius below 1 µm can support several kilowatts of heating power. A typical case with $R = 1.5\ \mu m$ can reach $Q_{h,60K} = 14.1\ kW.cm^{-2}$ which is around one order of magnitude higher than a conventional cooling system. These results demonstrate that, though our TCMC design might be more challenging to make experimentally than existing solutions, it can provide a leap in performance.

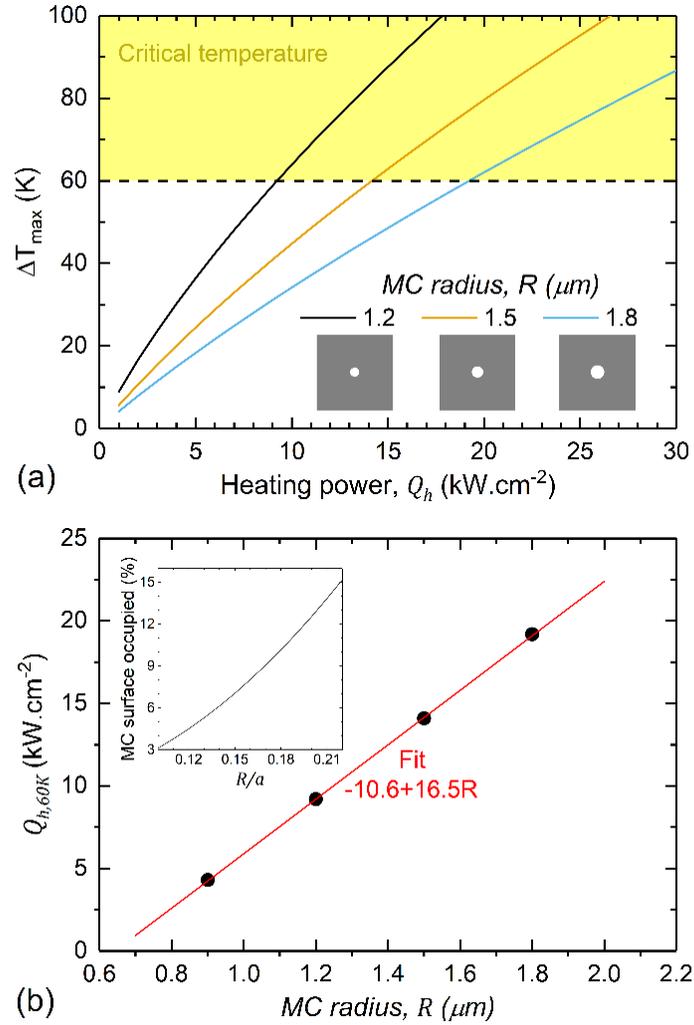

**Fig. 4.** (a) Impact of heating power on the maximum temperature increase with a critical temperature set at 60 K. (b) Linear relationship between the power supported at $\Delta T_{max} = 60\ K$ and R. Inset shows the surface occupation ratio of the MCs on the chip.

### 3.3. Impact of inlet pressure

The inlet pressure generates the forced convection that makes the fluid circulate in the MCs. It causes mechanical stress to the chip that should be considered during the structural design of the chip. Though we do not conduct structural mechanic simulations, we investigate the impact of the inlet pressure on the thermal properties here. $P_{in}$ is varied from 0.1 MPa to 2 MPa for three layers with $a = 10~\mu m$, and $Q_h = 10~kW.cm^{-2}$. Fig. 5(a) shows that $\Delta T_{max}$ decreases when $P_{in}$ increases, as expected because the increased water velocity carries away the heat more rapidly. However, the decrease slows as $P_{in}$ increases, and we can see that applying a pressure beyond 1 MPa only provides minor improvements. The MC radius R has a more significant impact on $\Delta T_{max}$ than $P_{in}$. Therefore, R should be considered a priority to limit the chip temperature.

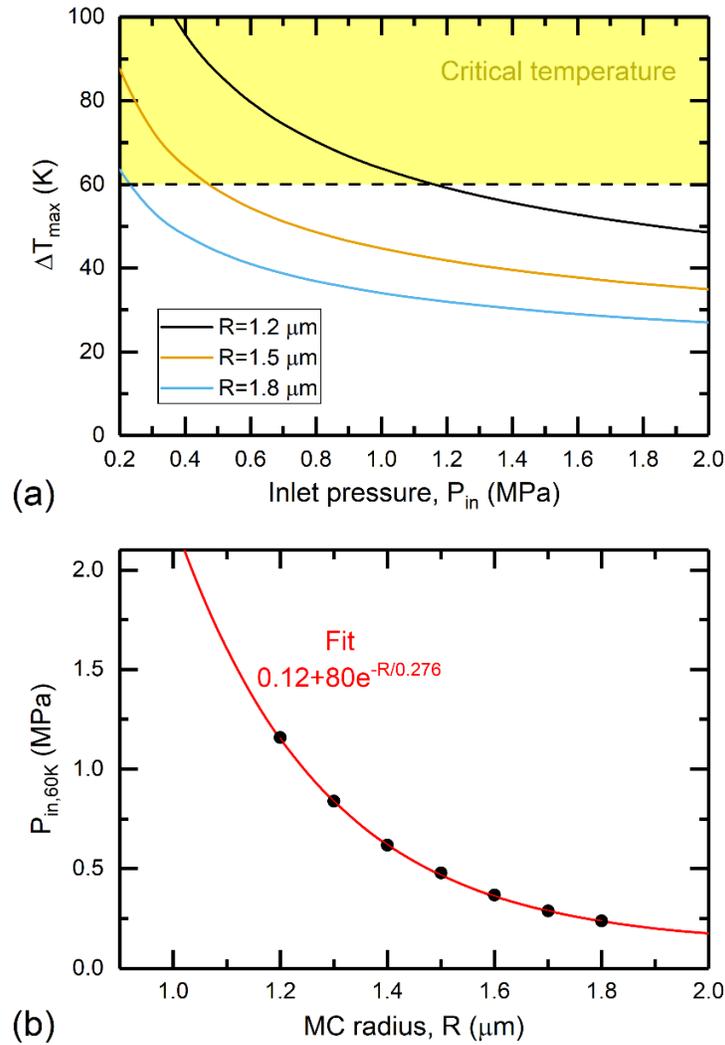

**Fig. 5.** (a) Impact of the inlet pressure on $\Delta T_{max}$. (b) Minimum inlet pressure required to get $\Delta T_{max} < 60~K$ and exponential fit.

In Fig. 5(b), we use the data in Fig. 5(a) to calculate the minimum inlet pressure $P_{in,60K}$ to keep $\Delta T_{max} < 60\ K$. It is found that $P_{in,60K}$ decreases exponentially with the MC radius following a law $P_{in,min} = 0.12 + 80 \exp(-R/0.276)$. Therefore, a MC radius of 1.5 µm with an inlet pressure as low as 0.48 MPa could sustain a heating power of $10\ kW.cm^{-2}$. This inlet pressure is reasonable in conventional MCs [46]. However, this pressure is applied perpendicularly to the surface of the chip in our TCMC design, which implies a large force is applied to the chip. Eventually, supporting beams could be added to the chip to increase its mechanical resistance to the pressure (see Appendix C). The exponential law also indicates that $P_{in}$ increases rapidly above 1 MPa when the radius is below 1.2 µm. Therefore, reducing the MC radius to save space for electronics comes at the cost of making the chip able to support very high pressures.

### 3.4. Impact of the number of layers

Up to now, we focused on a three-layer chip. Here we vary the number of layers $N$ from 1 to 5, in the case of $a = 10\mu m$, $R = 1.5\ \mu m$, $Q_h = 10\ kW.cm^{-2}$, and $P_{in} = 1 MPa$. We remind that the thickness of the chip, which also corresponds to the length of the MC, always remains constant at $L_{MC} = 37.5\ \mu m$. The supporting sub-layer and separating sub-layer have their thicknesses adjusted to keep $L_{MC}$ constant (see Fig. 1). The heating sub-layer also keeps a constant thickness $\delta = 1\ \mu m$. The heating power $Q_h$ is divided equally among the $n$ heating sub-layers. Consequently, increasing $N$ means that $Q_h$ is distributed more homogeneously in the volume of the chip.

Fig. 6(a) shows that the number of layers has no significant impact on $\Delta T_{max}$, and therefore, on the cooling performance of our TCMC design. The main reason is that our MC is in contact with all the sub-layers, providing cooling to each of them. Figs. 6(b), 6(c), and 6(d) show the temperature distribution for chips with one layer, two layers, and four layers, respectively. The temperature scale was normalized to be identical for the three plots. We can see temperature jumps between the layers in the solid domain because of the separating sub-layer that has poor thermal conductivity. However, the fluid temperature distribution remains identical in all cases. Furthermore, we found that the water mass flux and average outlet temperature rise also remain unchanged with the number of layers. Therefore, the cooling performance is limited by the capacity of the water to take away the heat while moving up the MC, independently of the local temperature in the solid.

Interestingly, even for just one layer, the cooling performance is superior to cooling by circulating water under the chip. The reason is that the cooling surface area by our TCMC design is bigger than the underneath surface. Indeed, from our unit cell, the underneath surface area is $a^2$ while the MC surface area is $2\pi Rh$. If we take $a = 10\ \mu m$ and $R = 1.5\ \mu m$, as we did previously, the TCMC provides a cooling surface area 3.5 times bigger than the underneath area (see supporting information). Therefore, TCMCs could not only serve to cool down 3D ICs but also for 2D chips.

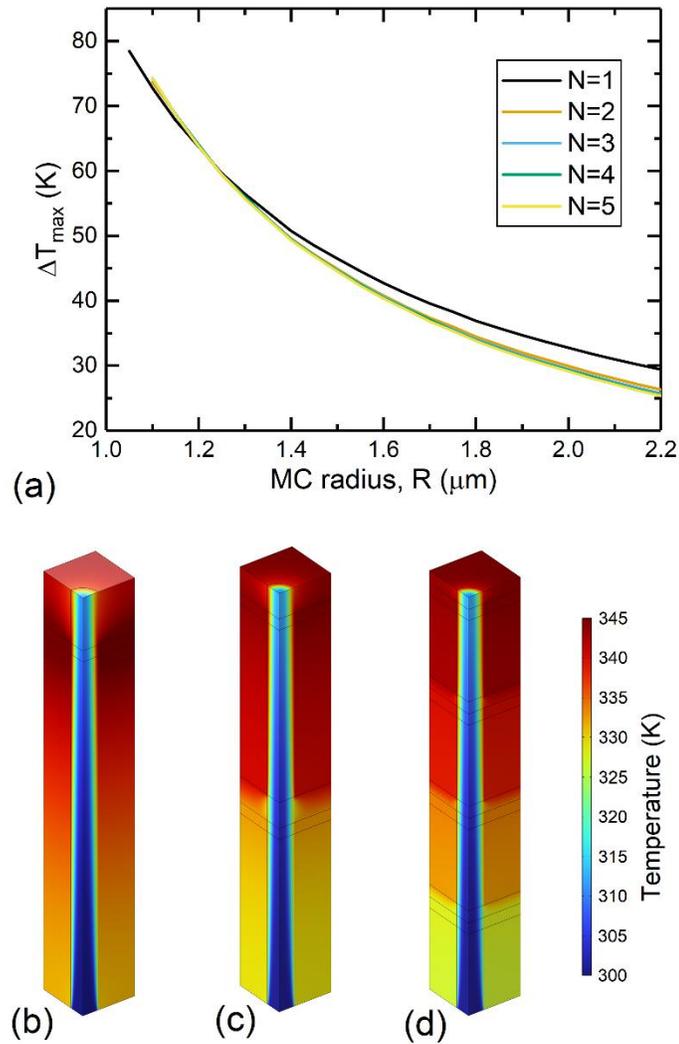

**Fig. 6.** (a) $\Delta T_{max}$ for different number of layers from 1 to 5. Temperature distribution in 1 layer (a), two layers (b), and four layers (d).

### 3.5. Manufacturing possibility

The new cooling system introduced here by the TCMCs is only studied numerically to establish its excellent performance and show that it is worth the challenge of realizing experimentally. In our TCMC design, both the 3D IC and the cooling system are embedded together, so the IC design must incorporate the microchannels. This is a paradigm shift compared to traditional chip manufacturing, where the IC design and cooling system are treated independently and even by different manufacturers. Manufacturing 3D ICs with TCMCs will require significant development to overcome several difficulties.

First, all circuit components must avoid the location where the microchannels pass, which necessitates rethinking the IC architecture. It should be noticed that the microchannels represent only 7% of the surface (with $a = 10\ \mu m$ and $R = 1.5\ \mu m$), as can be seen in the inset of Fig. 5(b). If a safety distance of 0.5 μm from the side of the microchannel is taken, 87% of the surface is still available for the components. Second, the water must flow through the microchannels without leaking into the IC. The layers must be water-tight bonded, or the surface of the microchannels could be coated with a material, creating a sheath. Third, the chip must withstand the hydraulic pressure. Supporting beams could be added to rigidify the chip as proposed in Appendix C, though it would further increase the fabrication complexity. Fourth, besides the chip, the outer part of the cooling system should also be designed to allow water to go through without leakage. A system of sealed electrical connections between the chip and the support board would have to be developed. Finally, beyond the technical issues, the cost of the TCMC design is expected to be high because of its complexity. However, all these difficulties should be put into perspective compared to the advantages.

The main positive aspect of the TCMC design is its breakthrough cooling performance, which can dissipate ten times more heat than existing cooling systems. To our knowledge, no other design can support heat powers above $10^4$ W.cm$^{-2}$. Another important feature is that the water can directly cool each layer of the 3D IC because TCMCs cross perpendicularly the whole chip. Therefore, TCMCs can allow the stacking of many layers and still provide efficient cooling for each. Also, the short length of the MCs provides a high flux of water with acceptable pressure drop. Finally, the risk of hot spots is almost inexistent because the MC array makes each IC component close to the coolant. These advantages can be worth further investigation to overcome the technical difficulties related to manufacturing.

## 4. Conclusions

In conclusion, we have designed a new cooling system for 3D ICs called *through-chip microchannels* (TCMCs) that is made of an array of MCs drilled perpendicularly to the layers through the entire chip. Our CFD simulations demonstrate that the TCMC design can directly bring considerable cooling power to the core of multilayer ICs. This study establishes the great potential of TCMCs to enable vertical 3D ICs that integrate many layers for dense, highly functional, and high-power electronics. TCMC design intertwines the IC and the liquid cooling system, which presents challenges for manufacturing. However, the prospect of obtaining cooling powers above $10 \ kW.cm^{-2}$ can represent a good incentive to support the experimental development of TCMCs. The major results are summarized as follows:

(1) The pitch and radius of the array of MCs were optimized to maintain the temperature rise $\Delta T_{max} < 60 \ K$. A pitch $a = 10 \ \mu m$ provides the best cooling performance of $14.1 \ kW.cm^{-2}$ with a radius $R = 1.5 \ \mu m$ and inlet pressure $P_{in}=1$ MPa. This small radius occupies only 7% of the surface of the chip to leave room for IC design.
(2) The maximum heating power increases linearly with the MC radius and can exceed $20 \ kW.cm^{-2}$ when $R > 1.8 \ \mu m$. This relation provides scalability between the surface available for electronic components and the heat power that must be dissipated depending on different applications.
(3) An exponential relationship between the inlet pressure and the MC radius is established. This relation can be used to minimize $P_{in}$ given the geometrical requirements to reduce the mechanical strain on the chip. E.g., a device that must dissipate $10 \ kW.cm^{-2}$ with $a = 10 \ \mu m$ and $R = 1.5 \ \mu m$ can use a pressure of 0.48 MPa.
(4) The cooling performance remains identical when increasing the number of layers with a fixed chip's length. Therefore, it is possible to stack many layers and create a chip with a very high electronic density by thinning the isolation between the layers.

**Declaration of Competing Interest**

The authors declare that they have no known competing financial interests or personal relationships that could have appeared to influence the work reported in this paper.

**Credit authorship**

**Lihong Ao**: Investigation, Methodology, Validation, Writing-Original draft. **Aymeric Ramiere**: Formal analysis, Supervision, Writing, Funding acquisition.

**Data availability**

Data supporting this work are available upon reasonable request to the authors.

**Acknowledgments**

This work was supported by the National Natural Science Foundation of China (No. 12050410254), and the Natural Science Foundation of Guangdong Province (No. 2023A1515011889).

**Appendix A: Error verification and validation**

An error verification as a function of the number of cells in the mesh of the geometry was conducted, as shown in Fig. 7(a). The geometry of our design being essentially parallelepiped, a mapped mesh was chosen in the vertical direction along the z-axis. Perpendicular to the z-axis, quadratic mesh was applied with refinement at the solid/fluid interface. The results show that the average temperature error is below 1 K for $5 \times 10^5$ cells. In our simulations, we used $7 \times 10^5$ which provided an error below 0.4 K.

To validate our results, we compare our simulations of the water velocity at the center of the MC to the maximum water velocity predicted by the well-established Poiseuille flow theory. In the case of a laminar flow of an incompressible fluid with non-slip wall conditions, the theoretical velocity at the center of the MC is given by $u_{max} = P_{in}R^2/(4\mu L_{MC})$. The geometry is taken with $R = 1.5 \ \mu m$ and $L_{MC} = 37.5 \ \mu m$. We take an average dynamic viscosity $\mu = 0.57 \ mPa.s^{-1}$ corresponding to the water at 320 K. Fig. 7(b) shows excellent agreement between the simulations and the theory. A slight deviation is observed due to the Maxwell slip boundary conditions $\alpha_v = 0.9$ in our simulations while the theory is for $\alpha_v = 1$. Fig 7(b) confirms that our simulations provide reasonable results and validate our method.

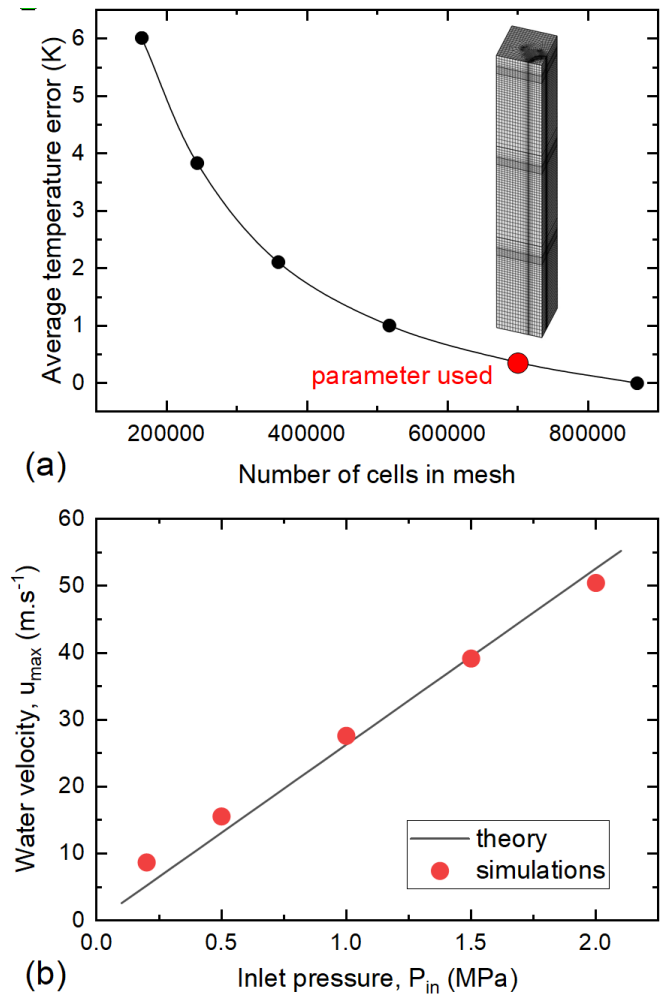

**Fig. 7.** (a) Average temperature error as a function of the number of cells. The inset shows an example of mesh used for the simulations. (b) validation of the results by comparing the water velocity to the Poiseuille flow theory.

**Appendix B: Thermophysical properties and frictional power**

Table 1 contains the thermophysical properties used in the CFD simulations. Si was used for the basis and heating sub-layers. $SiO_2$ was used for the insulation layers. All other quantities concern the water fluid circulating in the microchannel (MC).

**Table 1:** thermophysical expressions used in the CFD simulations.

| quantity | expression | unit |
|---|---|---|
| $\kappa_{Si}$ | 130 | $W/(m \cdot K)$ |
| $\kappa_{SiO_2}$ | $-0.98547 + 0.01821T - 5.2905 \times 10^{-5}T^2 + 7.5525 \times 10^{-8}T^3 - 5.0081 \times 10^{-11}T^4 + 1.3113 \times 10^{-14}T^5$ | $W/(m \cdot K)$ |
| $\kappa_f$ | $-0.86908 + 0.008949T - 1.5836 \times 10^{-5}T^2 + 7.9754 * 10^{-9}T^3$ | $W/(m \cdot K)$ |
| $\rho_f$ | $1.0335 \times 10^{-5}T^3 - 0.013395T^2 + 4.96929T + 432.257$ | $kg/m^3$ |
| $C_p$ | $12010.1471 - 80.4073T + 0.3099T^2 - 5.38187 \times 10^{-4}T^3 + 3.62537 \times 10^{-7}T^4$ | $J/(kg \cdot K)$ |
| $\mu_f$ | $1.37996 - 0.021224T + 1.36046 \times 10^{-4}T^2 - 4.6454 \times 10^{-7}T^3 + 8.90427 \times 10^{-7}T^4 - 9.07907 \times 10^{-13}T^5 + 3.84573 \times 10^{-16}T^6$ | $Pa \cdot s$ |

The frictional power was simulated independently as a function of the inlet pressure and the MC radius. The simulation results are shown with the blue surface in Fig. 8. This curve was then fitted with a polynomial function following the expression:

$$Q_{fric} = \frac{2.12}{a^2} P_{in}^{2.146}(0.8353R + 0.195065R^2 + 0.50322R^{4.368})\left(\frac{0.6032}{L_{MC}^2} + \frac{1.73}{L_{MC}}\right) \quad (14)$$

In the cases shown in our paper, $L_{MC} = 37.5\mu m$, so that Eq. 14 becomes:

$$Q_{fric} = 0.09871 P_{in}^{2.146}(0.8353R + 0.195065R^2 + 0.50322R^{4.368}) \quad (15)$$

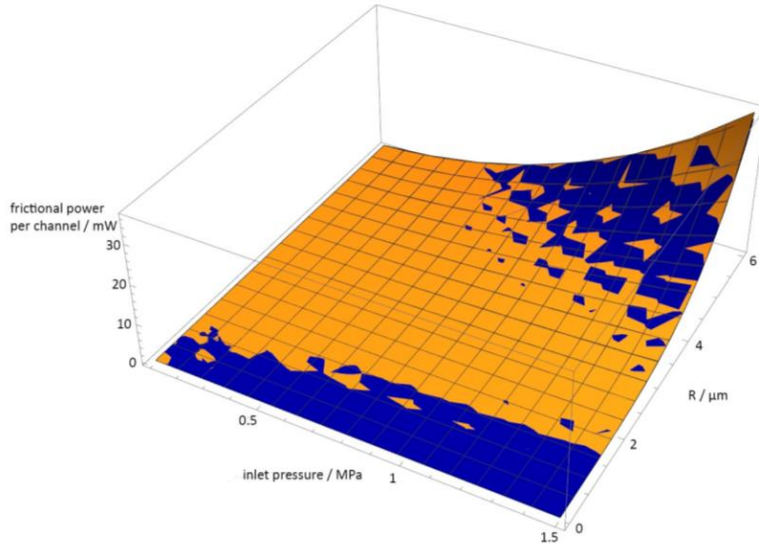

**Fig. 8**. Simulated $Q_{fric}$ (blue surface) and polynomial fit (orange surface).

**Appendix C: Supporting beams for high pressure**

In our simulations, we focus on the thermal aspect of the design and ignore the mechanical aspect. However, applying high pressures tangentially to the chip will cause significant mechanical stress that could eventually break the chip. To answer this concern, we propose in Fig. 9 a more complete chip design that includes supporting beams above the last layer and aims at rigidifying the chip to help it sustain high pressure. This is an indicative design where the thickness and eight of the supporting beams would need to be determined.

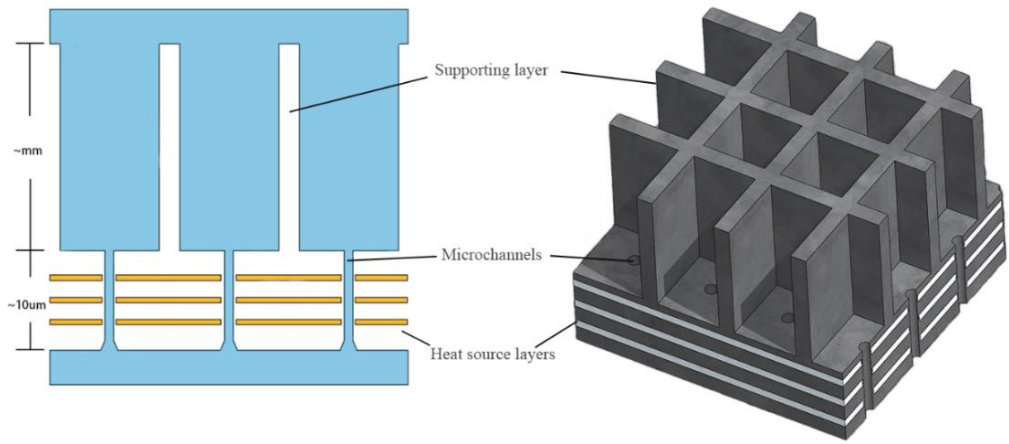

**Fig. 9.** Example of supporting beams to improve mechanical strength.

**Appendix D: Cooling surface ratio**

The cooling performance is related to the surface in contact with the fluid. Here, we compare the surface of the MC, noted $S_{MC}$, to the surface under the unit cell, noted $S_U$. The ratio of these two quantities is given by:

$$\frac{S_{MC}}{S_U} = \frac{2\pi R L_{MC}}{a^2}$$

In our case, $L_{MC} = 37.5 \ \mu m$ is fixed. Fig. 10 shows that in almost all situations $S_{MC}/S_u > 1$. Only in the bottom right corner, when $R/a$ is very small and $a$ is big, do we get a ratio below 1 where the area under the chip is higher than the area of the MC. However, the conditions where $S_{MC}/S_u < 1$ lead to $\Delta T_{max} > 73 \ K$, which were eliminated from our results because the water boiling temperature is exceeded. In all other cases, the area of the MC is bigger than the area under the chip.

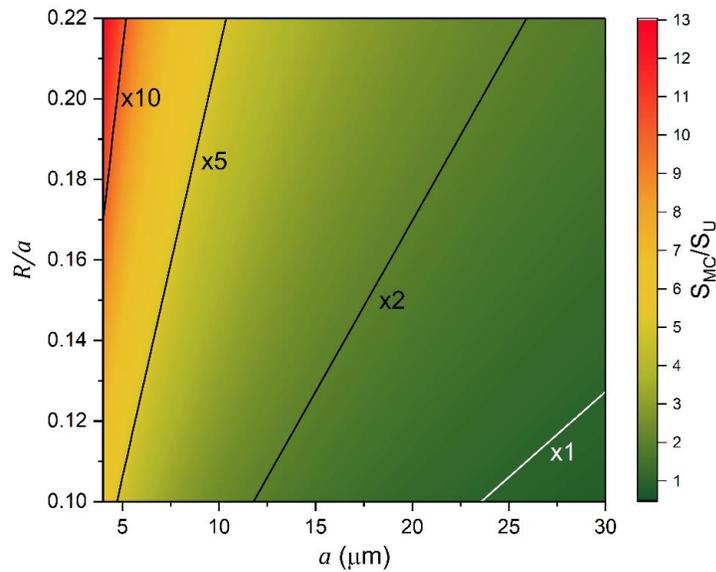

**Fig. 10.** Map of the MC surface over the surface under the unit cell $S_{MC}/S_u$.